\documentclass[final,1p,times]{elsarticle}
\usepackage{amssymb}
\newcommand{\ket}[1]{\mbox{$ | #1 \rangle $}}
\newcommand{\bra}[1]{\mbox{$ \langle #1 | $}}
\begin{document}

\begin{frontmatter}

\title{Semiquantum key distribution using entangled states}


\author{Jian Wang\fnref{fn1}}
\author{Sheng Zhang}
\author{Quan Zhang}
\author{Chao-Jing Tang}

\address{School of Electronic Science and Engineering, National University of Defense Technology, Changsha 410073, China}
\fntext[fn1]{Tel. No.: 86 0731 84575706, E-mail address: jwang@nudt.edu.cn}

\begin{abstract}
Recently, Boyer et al. presented a novel semiquantum key distribution protocol
[M. Boyer, D. Kenigsberg, and T. Mor, Phys. Rev. Lett. 99, 140501 (2007)],
by using four quantum states, each of which is randomly prepared by $Z$ basis or $X$ basis.
Here we present a semiquantum key distribution protocol by using entangled states
in which quantum Alice shares a secret key with classical Bob.
We also show the protocol is secure against eavesdropping.
\end{abstract}


\end{frontmatter}

% \linenumbers

\section{Introduction}
\label{}
Quantum key distribution (QKD) is one of the most promising
applications of quantum information science. The goal of QKD is to
allow two legitimate parties, Alice and Bob, to generate a secret
key over a long distance, in the presence of an eavesdropper, Eve,
who interferes with the signals. The security of QKD is based on the
fundamental laws of physics. Since the BB84
protocol\cite{BB84}, the first QKD scheme, was published, many
variations on QKD have been subsequently proposed. They can be
roughly classified into ``prepare and measure'' protocols, such as
BB84, B92\cite{B92}, the three-state protocol\cite{HA00}, the
six-state protocol\cite{B98} and ``entanglement based'' protocols,
such as E91\cite{E91}, BBM92\cite{BBM92}. There have been efforts made to
set a security proof based on entanglement for the both
classes\cite{CLL04}.

QKD protocols claim that both the parties are quantum.
That is to say, both parties should perform quantum operation on their qubits.
Boyer et al. considered a new problem that in the QKD protocol
if one party is quantum and the other has only classical capabilities,
whether the protocol could achieve to distribute a key between the parties securely.
They then presented a novel semiquantum key distribution protocol\cite{BKM07}
(hereafter called BKM07 protocol), in which one party (Alice) is quantum and the other (Bob) is classical,
and proved that the protocol is completely robust against an eavesdropping attempt.
Similar to BB84, in their protocol, Alice prepares a random qubit
in the computational ($Z$) basis \{\ket{0},\ket{1}\} or
Hadamard ($X$) basis \{\ket{+},\ket{-}\} and sends it to Bob.
They call the computational basis classical
and use the classical notion \{0,1\} to describe the two quantum states
\{\ket{0},\ket{1}\}. Bob can access a segment of the quantum channel,
but he is restricted to performing some classical operations. Different from BB84,
he can only measure the transmission qubits in the classical
\{0,1\} basis, prepare a qubit in the classical basis and send it,
reflect it to Alice directly or reorder the qubits (for example, by using different delay lines).
As Bob can only deal with qubits in the classical basis,
he is regarded as a classical party.

To ensure the security of BKM07 protocol, the communication parties should keep Eve from knowing
which is a particle that Bob measured in the computational basis and
which is a particle that Bob reflected to Alice directly.
Actually, BKM07 protocol is similar to BB84 protocol if we regard Bob as the initiator in BKM07 protocol.
Bob selects randomly either to measure the receiving qubit and resend it
or reflect it directly, which looks like Bob prepares randomly
the four quantum states \{\ket{0},\ket{1},\ket{+},\ket{-}\}.
Boyer et al. showed a different semiquantum key distribution protocol\cite{BGKM09}
based on randomization (hereafter called BGKM09-Randomization-based protocol).
Different from BKM07 protocol, Bob can reorder the particles besides measuring
and preparing a qubit in the classical basis in BGKM09-Randomization-based protocol.
In order to avoid Eve's acquiring Bob's
operation information on each receiving qubit, Bob reorders randomly the reflected
qubits. Furthermore, they proved the robustness of the protocol
in much more general scenario. Zou et al. presented five different semiquantum
key distribution protocols\cite{ZOU09} in which Alice sends three quantum states, two
quantum states and one quantum state, respectively. Li et al. proposed
two semiquantum secret sharing protocols\cite{LI10} by using maximally entangled
Greenberger-Horne-Zeilinger states in which quantum Alice shares a secret
with two classical parties, Bob and Charlie.

In this Letter, we present a semiquantum key distribution protocol by using entangled states.
We follow the descriptions about classical in Ref.\cite{BKM07}.
Quantum Alice can prepare two-particle entangled states and measure the particles
in Bell basis or computational basis. Classical Bob is restricted to
measuring, preparing, reordering, or sending quantum states only in the fixed orthogonal quantum basis set
\{\ket{0},\ket{1}\}. We show that Eve's eavesdropping would inevitably disturb
the transmission quantum states and the communication parties could detect
Eve's attack.

\section{Description of the protocol}

%% \section{Description of the protocol}
We then describe the semiquantum key distribution protocol using entangled states
in detail. Suppose Alice wants to share a secret key with classical Bob.
The protocol comprises the following steps.

(1) Alice prepares $N$ Bell states, each of which is in the state
\begin{eqnarray}
& &\ket{\phi^+}=\frac{1}{\sqrt{2}}(\ket{00}+\ket{11})_{AB},
\end{eqnarray}
where A, B represent the two particles of each state. Alice
takes one particle from each state to form an ordered partner
particle sequence $[B_1,B_2,\cdots,B_N]$, called $B$
particle sequence. She then send the $N$ particles $[B_1,B_2,\cdots,B_N]$ to Bob.

(2) For each particle arriving, Bob selects randomly either to measure it
in the computational basis (SIFT it) or to reflect it to Alice directly (CTRL it).
Bob reorders randomly the $N$ particles and generates a rearranged particle sequence
$[B'_1,B'_2,\cdots,B'_N]$, called $B'$ particle sequence.
He then sends $B'$ particle sequence to Alice. The
order of the rearranged particle sequence is completely secret to others but Bob
himself, which ensures the security of the present scheme.

(3) Alice stores the receiving particles in quantum memory and
informs Bob that she has received the $N$ particles.
Bob publishes which particles he chose to CTRL and the secret rearranged order of the particle sequence.

(4) For each of the CTRL particles (those were reflected directly),
Alice performs Bell basis measurement on the reflected particle B and particle A she owned.
If there is no eavesdropping, Alice's measurement result should be \ket{\phi^+}.
Alice analyzes the error rate on the CTRL particles according to her measurement results.
If the error rate is higher than the threshold they preset,
they abort the protocol. We call this step the first eavesdropping check.

(5) For each of the SIFT particles (those were measured in the computational basis),
Alice performs Z basis measurement on particle A and particle B.
She chooses randomly a sufficiently large subset from the measurement results for eavesdropping check.
She announces which are the chosen particles. Bob then publishes his measurement results.
Alice checks the error rate and if it is below the threshold they preset, they
proceed to execute the next step. Otherwise they abort the protocol.
We call this step the second eavesdropping check.

(6) Alice announces error correction code (ECC) and privacy amplification (PA) data.
Alice and Bob utilize the ECC and PA to extract the final key.

Obviously, the protocol can also be realized by using some other entangled states, such as $\ket{\phi^-}=\frac{1}{\sqrt{2}}(\ket{00}-\ket{11})_{AB}$
, $\ket{\psi^+}=\frac{1}{\sqrt{2}}(\ket{01}+\ket{10})_{AB}$ and
$\ket{\psi^-}=\frac{1}{\sqrt{2}}(\ket{01}-\ket{10})_{AB}$.
Similar to BKM07 protocol, Bob can make use of a different sifting method called measure-resend at step (2).
We only need to modify step (2)-(3) of the previous protocol. The modified steps are as follows:

(2') For each particle arriving, Bob selects randomly either to measure it
in the computational basis and resend it to Alice after measurement (SIFT it)
or to reflect it to Alice directly (CTRL it).

(3') Alice stores the receiving particles in quantum memory and
informs Bob that she has received the $N$ particles.
Bob publishes which particles he chose to SIFT and which particles he reflected directly to Alice.

We call the modified protocol measure-resend protocol. In the measure-resend protocol,
Bob only needs to perform computational basis measurement on the particles and reflect the particles directly.

\section{The security for the protocol}
The key of the security of the protocol is to keep Eve from knowing
which particles are CTRL particles and which ones are SIFT particles.
If Eve can distinguish CTRL and SIFT particles properly, she can easily
achieve the secret key by performing computational basis measurements on
the SIFT particles. At step (2) of the protocol, Bob reorders the particles
so that no one could distinguish CTRL and SIFT particles properly except Bob.
However, Eve can distinguish Bob's operation properly with a certain probability
in the measure-resend protocol. Of course,
Eve's eavesdropping will be detected by communication parties during the second eavesdropping check.
Suppose Eve intercepts particle B at step (1) and prepares a Bell state
$\ket{\phi^+}_{EE'}=\frac{1}{\sqrt{2}}(\ket{00}+\ket{11})_{EE'}$. She sends particle $E'$
to Bob instead of particle B and keeps particle $E$.
She reflects particle B directly to Alice without disturbing it. According to step (2'),
Bob chooses randomly either to SIFT it or to CTRL it. If Bob chooses to SIFT,
$\ket{\phi^+}_{EE'}$ will collapse to
$\ket{00}_{EE'}=\frac{1}{\sqrt{2}}(\ket{\phi^+}+\ket{\phi^-})_{EE'}$ or
$\ket{11}_{EE'}=\frac{1}{\sqrt{2}}(\ket{\phi^+}-\ket{\phi^-})_{EE'}$ each with
probability of 1/2. If Bob chooses to CTRL, $\ket{\phi^+}_{EE'}$ keeps unchanged.
After Bob sending particle $E'$ to Alice, Eve intercepts particle $E'$ and
performs Bell basis measurement on particle $E$ and $E'$. If she obtains
\ket{\phi^-}, she can ascertain that Bob has chosen to SIFT. If Bob chooses to SIFT
or CTRL each with probability of 1/2, Eve can acquire his operation information
properly with probability of 1/4. The communication parties cannot detect
the exist of Eve during the first eavesdropping check. However,
Eve's attack cannot escape from the second eavesdropping check.
We'd best not let Eve have a chance of distinguishing CTRL and SIFT particles properly.
For the protocol using order rearrangement, this trick of Eve's
will inevitably fail.

According to Stinespring dilation theorem\cite{Stinespring}, the attack
of an eavesdropper Eve can be realized by a unitary operation
$\hat{E}$ on a large Hilbert space, $H_{AB}\otimes H_{E}$. Then the
state of Alice, Bob and Eve is
\begin{eqnarray}
\ket{\Phi}=\sum_{a,b\in\{0,1\}}\ket{\varepsilon_{a,b}}\ket{a}\ket{b},
\end{eqnarray}
where \ket{\varepsilon} denotes Eve's probe state. \ket{a} and
\ket{b} are states shared by Alice and Bob. The condition on the
states of Eve's probe is
\begin{eqnarray}
\sum_{a,b\in\{0,1\}}\bra{\varepsilon_{a,b}}\;
\varepsilon_{a,b}\rangle=1.
\end{eqnarray}
We can describe Eve's attack on the system as
\begin{eqnarray}
& &\hat{E}\ket{0,\varepsilon}=\alpha\ket{0,\varepsilon_{00}}+\beta\ket{1,\varepsilon_{01}},\\
&
&\hat{E}\ket{1,\varepsilon}=\beta'\ket{0,\varepsilon_{10}}+\alpha'\ket{1,\varepsilon_{11}},
\end{eqnarray}
where $|\alpha|^2+|\beta|^2=1, \beta'|^2+|\alpha'|^2|=1$.
Eve can eavesdrop particle B at step (1) and the state of composite
system will be
\begin{eqnarray}
\ket{\Phi}&=&\frac{1}{\sqrt{2}}[\ket{0}_A(\alpha\ket{0,\varepsilon_{00}}+\beta\ket{1,\varepsilon_{01}})_{BE}
+\ket{1}_A(\beta'\ket{0,\varepsilon_{10}}+\alpha'\ket{1,\varepsilon_{11}})_{BE}]
\nonumber\\
&=&\frac{1}{\sqrt{2}}[\alpha\ket{0,0,\varepsilon_{00}}+\beta\ket{0,1,\varepsilon_{01}}
+\beta'\ket{1,0,\varepsilon_{10}}+\alpha'\ket{1,1,\varepsilon_{11}}]_{ABE}\nonumber\\
&=&\frac{1}{\sqrt{2}}[(\alpha\ket{0,\varepsilon_{00}}+\beta'\ket{1,\varepsilon_{10}})_{AE}\ket{0}_B
+(\beta\ket{0,\varepsilon_{01}}+\alpha'\ket{1,\varepsilon_{11}})_{AE}\ket{1}_B].\nonumber\\
\end{eqnarray}
If Bob chooses to SIFT, \ket{\Phi} will collapse to $(\alpha\ket{0,\varepsilon_{00}}+\beta'\ket{1,\varepsilon_{10}})_{AE}\ket{0}_B$
or $(\beta\ket{0,\varepsilon_{01}}+\alpha'\ket{1,\varepsilon_{11}})_{AE}\ket{1}_B$ each with probability of 1/2.
In the course of  the communication parties' checking the errors on the SIFT particles at step (5),
the error rate introduced by Eve is $e=|\beta|^2=|\beta'|^2=1-|\alpha|^2=1-|\alpha'|^2$.
If Bob chooses to CTRL, particle B will be reflected directly. Eve can intercept particle B again and
perform another unitary operation $\hat{E'}$ on \ket{\Phi},
the state of the system becomes
\begin{eqnarray}
\ket{\Phi'}&=&\frac{1}{\sqrt{2}}[\gamma\ket{0,0,\varepsilon'_{00}}+\delta\ket{0,1,\varepsilon'_{01}}
+\delta'\ket{1,0,\varepsilon'_{10}}+\gamma'\ket{1,1,\varepsilon'_{11}}]_{ABE}\nonumber\\
&=&\frac{1}{2}[\gamma(\ket{\phi^+}+\ket{\phi^-})_{AB}\ket{\varepsilon'_{00}}_E
+\delta(\ket{\psi^+}+\ket{\psi^-})_{AB}\ket{\varepsilon'_{01}}_E\nonumber\\
&+&\delta'(\ket{\psi^+}-\ket{\psi^-})_{AB}\ket{\varepsilon'_{10}}_E
+\gamma'(\ket{\phi^+}-\ket{\phi^-})_{AB}\ket{\varepsilon'_{11}}]_E.
\end{eqnarray}
When the communication parties check the errors on the CTRL particles at step (4),
the error rate introduced by Eve is $e=1-(|\gamma|^2+|\gamma'|^2)/4$.
If Eve attempts to obtain information, she would inevitably induce some
errors that legitimate parties could notice.

To explain the error rate introduced by Eve, we take an example for individual attack strategy.
Suppose Eve intercepts the particles in $B$ particle sequence transmitted to Bob at step (1)
and makes measurements on them. She then resends particles to Bob according to
her measurement results. In this way, if Eve measures particle B in the $Z$ basis,
\ket{\phi^+} will collapse to $\ket{00}_{AB}$ or $\ket{11}_{AB}$. If Eve's result is \ket{0} (\ket{1}), she
sends a particle in the state \ket{0} (\ket{1}) to Bob. According to the protocol,
Bob chooses randomly either to SIFT it or to CTRL it and reorders the particles.
Because $\ket{00}=\frac{1}{\sqrt{2}}(\ket{\phi^+}+\ket{\phi^-})$
($\ket{11}=\frac{1}{\sqrt{2}}(\ket{\phi^+}-\ket{\phi^-})$), during the first eavesdropping check,
Alice performs Bell basis measurement on particle A and B, and obtains \ket{\phi^+} or \ket{\phi^-}
each with probability of 1/2. Thus the error rate introduced by Eve is 50\%.
Suppose Eve intercepts particle B at step (1) and uses it and her own ancillary particle
in the state $\ket{0}$ to do a CNOT operation (particle B is the controller, Eve's ancillary particle is the target).
Eve then resends particle B to Bob. The state of particle A, B and Eve's ancillary particle becomes
$\frac{1}{\sqrt{2}}(\ket{000}+\ket{111})_{ABE}$. At step (2),
Bob selects randomly either to SIFT or to CTRL and reorders the particles.
In view of $\frac{1}{\sqrt{2}}(\ket{000}+\ket{111})_{ABE}=\frac{1}{2}[(\ket{\phi^+}+\ket{\phi^-})\ket{0}+(\ket{\phi^+}-\ket{\phi^-})\ket{1}]$, Alice obtains \ket{\phi^+} or \ket{\phi^-} each with probability of 1/2 at step (4).
Thus the error rate introduced by Eve will reach 50\% and her eavesdropping
will be detected.

\section{Conclusion}
So far we have proposed a semiquantum key distribution protocol by using entangled states
and analyzed the security for the present protocol. In the protocol, quantum Alice can
share a secret key with classical Bob. Bob is restricted to measuring a particle in the classical basis,
preparing a particle in the classical basis, reflecting a particle directly, or reordering the particles.
Bob utilizes order rearrangement to keep Eve from
distinguishing which particles Bob chose to CTRL and which particles he chose to SIFT.
The communication parties make use of twice eavesdropping checks to ensure the security of
the protocol. Since the users only need to perform some classical operations on particles,
semiquantum key distribution can be realized at a low cost.
It also provides a good idea for building quantum key distribution network
(For example, we can build a Key Distribution Center which is quantum, but the users have only classical capabilities).
There are many difficulties to be dealt with when implementing semiquantum key distribution
in the practical scenario. We would like to explore these problems in future.

.

\section*{Acknowledgements}
This work is supported by the National Natural Science Foundation of
China under Grant No. 60872052.

%% References


\bibliographystyle{model1a-num-names}
\bibliography{mybib}

%%\bibitem{BB84} C. H. Bennett and G. Brassard, \textit{in Proceedings of IEEE
%%international Conference on Computers, Systems and signal
%%Processing, Bangalore, India} (IEEE, New York), pp. 175 - 179
%%(1984).
%%\bibitem{B92} C. H. Bennett, Phys. Rev. Lett. \textbf{68}, 3121 (1992).
%%\bibitem{ha00}H. Bechmann-Pasquinucci and A. Peres,  Phys. Rev. Lett.
%%\textbf{85}, 3313 (2000).
%%\bibitem{B98} D. Bruss, Phys. Rev. Lett. \textbf{81}, 3018 (1998).
%%\bibitem{E91} A. K. Ekert, Phys. Rev. Lett. \textbf{67}, 661 (1991).
%%\bibitem{BBM92} C. H. Bennett, G. Brassard and N. D. Mermin, Phys. Rev. Lett.
%%\textbf{68}, 557 (1992).
%%\bibitem{CLL04} M. Curty, M. Lewenstein, and N. L¡§utkenhaus,
%%Phys. Rev. Lett. \textbf{92}, 217903 (2004).
%%\bibitem{BKM07} M. Boyer, D. Kenigsberg, and T. Mor, Phys. Rev. Lett. \textbf{99},
%%140501 (2007).
%%\bibitem{BGKM09} M. Boyer, R. Gelles, D. Kenigsberg, and T. Mor, Phys. Rev. A \textbf{79},
%%032341 (2009).
%%\bibitem{ZOU09} X. Zou, D. Qiu, L. Li, L. Wu, and L. Li, Phys. Rev. A \textbf{79},
%%052312 (2009).
%%\bibitem{LI10} Q. Li, W. H. Chan, and D. Y. Long, Phys. Rev. A \textbf{82},
%%022303 (2010).
%%\bibitem{Stinespring} W. F. Stinespring, Proc. Am. Math. Soc. \textbf{6},
%%211 (1955).
%%\end{thebibliography}






% \end{thebibliography}

\end{document}